\documentclass[letterpaper, 10 pt, conference]{ieeeconf}  % Comment this line out if you need a4paper

\IEEEoverridecommandlockouts                              % This command is only needed if 
\usepackage{graphicx,xcolor} % for pdf, bitmapped graphics files
\usepackage{amsmath} % assumes amsmath package installed
\usepackage{amssymb}  % assumes amsmath package installed

\title{\LARGE \bf
A variational and symplectic framework for model-free control: preliminary results
}

\author{Loïc Michel
\thanks{Loïc Michel is with Nantes Universite-\'Ecole Centrale de Nantes-LS2N, UMR 6004 CNRS, Nantes, France
        {\tt\small loic.michel@ec-nantes.fr}}%    
}

\usepackage{amsmath,amssymb,amsfonts}

\newcommand{\diff}{\mathrm{d}}
\begin{document}

\maketitle
\thispagestyle{empty}
\pagestyle{empty}

%%%%%%%%%%%%%%%%%%%%%%%%%%%%%%%%%%%%%%%%%%%%%%%%%%%%%%%%%%%%%%%%%%%%%%%%%%%%%%%%
\begin{abstract}
The model-free control approach is an advanced control law that requires few information about the process to control. Since its introduction in 2008, numerous applications have been successfully considered, highlighting attractive robustness properties towards tracking efficiency and disturbance rejection.
In this work, a variational approach of the model-free control is proposed in order to extend its robustness capabilities. An adaptive formulation of the controller is proposed using the calculus of variations within a symplectic framework, that aims to consider the control law as an optimization problem toward the auto-tuning of its main key parameter. The proposed formulation provides a coupling between the model-free control law and a variational integrator to improve the robustness of the tracking towards process changes and emphasize closed-loop stabilization. Some illustrative examples are discussed to highlight the rightness of the proposed approach.
\end{abstract}

%%%%%%%%%%%%%%%%%%%%%%%%%%%%%%%%%%%%%%%%%%%%%%%%%%%%%%%%%%%%%%%%%%%%%%%%%%%%%%%%
\section{INTRODUCTION}

The model-free control methodology, proposed in \cite{Fliess}, has been designed to control { a priori} any "unknown" dynamical system and can be considered as an alternative to standard PI control \cite{FliessJoin_2021} as it does not need any prior knowledge of the plant to control (also referred as a "self-tuning regulator" in \cite{kumar}). Its usefulness in many situations, including compensating severe non-linearities and time-varying reference signals, has been demonstrated throughout successful applications in a wide range of engineering fields\footnote{See e.g. the references in \cite{Fliess,Bara,Hamiche} and the references therein for an overview of the applications.}, like for instance, glycemia control \cite{moog}, walking robot movement control \cite{tabuada}, local aerodynamic lift management \cite{wes-10}, wind energy conversion system \cite{wind_turbine}, multitasks pneumatic gripper \cite{gripper} and the control resource harvesting in a computing
grid \cite{computing}. The corresponding intelligent controllers are much easier to implement and to tune than standard PID controllers which are today the main tool in industrial control engineering (see, e.g., \cite{feedbacksystems}).

This algorithm assumes that the controlled dynamical system can be locally described by a polynomial model that is continuously updated, from which non-linearities of the process are compensated by the associated model-free controller. It is composed of the "model-free" part (tuned by a key parameter) and a usual PI controller.  
 In \cite{gedouin} \cite{rotella} \cite{automated_car}, the adaptation / estimation of the key tuning parameter regarding specific application cases have been addressed, hence highlighting some trends to improve the closed-loop behavior by adding dedicated adaptation laws with respect to the control problems.

In this paper, we propose to use variational principles in order to tune online the model-free control through a Lagrangian formulation within a symplectic framework.
The seminal works of J.E. Marsden (e.g. \cite{Abraham1967Foundations}) established the symplectic integrators \cite{WestPhD} \cite{Ruth} as a class of numerical methods specifically designed for solving Hamiltonian systems, which play a fundamental role mainly in mechanics, and molecular dynamics. These systems are governed by Hamilton's equations, which describe the evolution of a system in terms of its positions and momenta. One of the key properties of Hamiltonian systems is the preservation of the symplectic structure, that is a geometric feature that ensures the conservation of phase, space volume and other invariants like energy over the time. The variational calculus \cite{brunt}, associated to the discrete integration of a Lagrangian under symplectic geometry assumptions \cite{leok2010} \cite{Marsden} provides simple integration algorithms that can preserve physical properties of the system and good convergence properties, for which research is very active, providing for example algorithms for optimal control \cite{trelat}, Hamiltonian systems managed by symplectic numerical method \cite{Chen_symp}, optimization on manifolds \cite{optim_manifold}, nonlinear mechanical systems \cite{KOTYCZKA}, and multi-rate integrators \cite{variationalmultirateintegrators}.

The contribution of this work takes benefits of this kind of integrator to build an online optimization algorithm, hence allowing a self-tuning of the control law.
The proposed variational-based model-free control formulation aims at considering the tracking error as a Lagrangian to minimize, thus providing an additional dynamic towards the adaptation of the key tuning-parameter. Such variational-based dynamic would help to "absorb" transients induced by internal modifications of the controlled process.

The paper is structured as follow. Section II reviews the model-free control principles. Section III presents the Lagrangian formulation and derive the  variational form of the model-free control law.
Section IV illustrates the behavior of the proposed approach and discusses the results. Section V gives some concluding remarks and perspectives.

\section{Model-Free Control principles}

Full details on model-free control are given in \cite{Fliess}.   

\noindent
We consider a SISO dynamical system: 

\begin{equation}\label{eq:gen_sys}
\left\{ \begin{array}{l}
\dot{x} = f(x,u) \\
y = g(x)
\end{array} \right.
\end{equation}
\noindent
The unknown description of the dynamical system \eqref{eq:gen_sys} is replaced by an ultra-local model (\textit{i.e.} that approximates very locally the overall dynamics of the system):
\begin{equation} \label{mod}
\frac{\diff^{(n)} y}{\diff t^{(n) }} = {F} + \alpha u
\end{equation} %the notation ${y}^{ }$ designates the $n$th time-derivative of $y$;
\noindent where the control and output variables are respectively $u$ and $y$; the time-varying quantity {$F$} is estimated online and subsumes the unknown internal structure and the external disturbances.  %\textcolor{blue}{More precisely, as we are under-sampled, {$F$} at the time $k-1$ is estimated with respect to the error at the time $k$, and it is assumed that $F$ at time $k$ is close to {$F$} at time $k-1$.} 
The constant $\alpha \in \mathbb{R}^{*, +}$ is the key parameter that is chosen by the practitioner such that the time-derivative of ${y}$ and $\alpha u$ are of the same magnitude. Therefore, $\alpha$ does not need to be precisely estimated. Practically, the ultra-local model can be chosen of order $n = 1$.

\noindent
Equation \eqref{mod} is only valid during a short time lapse that must be continuously updated: it implies that ${F}$ is estimated online through the knowledge of the control output $u$ and the numerical differentiation of $y$. To numerically estimate the derivative of $y$, for example, discrete-time differentiator techniques can be used (see e.g. \cite{CEP_Rasool} for a survey). 

\textit{ Intelligent PI controllers}

The control law reads as the \textit{intelligent PI controller}, or \textit{i-PI} controller:
\begin{equation}\label{ip}
u = -  \frac{F}{\alpha} + \frac{1}{\alpha} \frac{\diff^{} y^{\ast}}{\diff t^{}}  + K_{{p}}  e + K_i \int e \, \diff \tau
\end{equation}

%where
%\begin{itemize}
%\item $y^\ast$ is the output reference trajectory;
%\item $e = y^\ast - y$ is the tracking error;
%\item $K_{\mathrm{p}}$ is a usual tuning gain.
%\end{itemize}
\noindent where $\alpha$ is the key parameter and $K_{{p}}, K_{{i}}$ are the usual tuning gain of the PI controller that have to be set by the user. The tracking error $e$ is defined by $y - y^{\ast}$.

\noindent
The i-PI controller \eqref{ip} is compensating the poorly known term ${F}$. Controlling the system therefore boils down to the control of an elementary pure integrator.

A recent study \cite{FliessJoin_2021} highlights that the ultra-local model \eqref{mod} can be chosen of the first order $(n = 1)$ in most applications and moreover, a reduced i-P controller can achieve better performances than i-PI control.

%The paper is structured as follows.

\section*{III. Lagrangian formulation}

In this section, a discretized Lagrangian formulation is derived from \eqref{mod} and \eqref{ip}.

\subsection*{Discrete i-P controller}

The controller is typically implemented in discrete time. Defining the sampling time $t_k = k h$ being the $k$th sampling  (with sampling period $h$) at which the control and output are updated.
Consider $n = 1$ and the approximation of the derivatives at the general discrete instant $k$:

\begin{equation}
\left. \frac{\diff y}{\diff t} \right|_{k} \approx  \frac{y_{k} - y_{k-1}}{h} 
\end{equation}

\begin{equation}
\left. \frac{\diff y^*}{\diff t} \right|_{q} \approx  \frac{y^*_{k} - y^*_{k-1}}{h} 
\end{equation}
\noindent
For the sake of simplicity, the $\diff / \diff t$ notation will be kept throughout all the developments.

\noindent
It is also assumed to use a pure proportional corrector $K_p$. Hence, an approximation of the ultra-local model  at $t = t_{k-1}$ reads:

\begin{equation}\label{eq:cont_1a}
\left. \frac{\diff^{ } y}{\diff t^{ }} \right|_{k-1} = F_{k-1} + \alpha u_{k-1}
\end{equation}

\begin{equation}\label{eq:cont_1b}
u_{k} = - \frac{F_{k-1}}{\alpha} +  \frac{1}{\alpha} \left. \frac{\diff^{ } y^*}{\diff t^{ }} \right|_k + K_p (y_k - y^*_k)
\end{equation}

\noindent
Substituting (\ref{eq:cont_1a}) into (\ref{eq:cont_1b}), we have:

\begin{equation}\label{eq:cont_1c}
u_{k} = u_{k-1} - \frac{1}{\alpha} \left\{\left. \frac{\diff^{ } y}{\diff t^{ }} \right|_{k-1} - \left. \frac{\diff^{ } y^*}{\diff t^{ }} \right|_k \right\} + K_p(y_k - y^*_k) 
\end{equation}
\noindent
where $K_p \in \mathbb{R}^{*, +}, \alpha \in \mathbb{R}^{*, +}$, which defines a $u_k$-recursive form associated to a time-derivative based anticipation of the proportional control.

\noindent
This definition is the "standard" discrete version to be implemented as a digital model-free controller.

\subsection*{Dynamical evolution of the tracking error $\varepsilon_k$}

Before presenting the optimization problem that consists in minimizing the Lagrangian, let us deduce from \eqref{eq:cont_1a}, \eqref{eq:cont_1b} and \eqref{eq:cont_1c} the evolution of the error:

\begin{equation}
u_{k} = - \frac{F_{k-1}}{\alpha} +  \frac{1}{\alpha} \left. \frac{\diff^{ } y^*}{\diff t^{ }} \right|_k + K_p (y_k - y^*_k)
\end{equation}

\begin{equation}
\alpha u_{k} = - F_{k-1} +  \left. \frac{\diff^{ } y^*}{\diff t^{ }} \right|_k + \alpha K_p (y_k - y^*_k)
\end{equation}
\noindent
and using the discrete ultra-local model  \eqref{eq:cont_1a}, one gets in the same manner at $t = t_{k}$:
\begin{equation}
\left. \frac{\diff^{ } y}{\diff t^{ }} \right|_{k} = F_{k} + \alpha u_k
\end{equation}

\noindent
Hence:
\begin{equation}
\left. \frac{\diff^{ } y}{\diff t^{ }} \right|_k - F_{k} = - F_{k-1} +  \left. \frac{\diff^{ } y^*}{\diff t^{ }} \right|_k + \alpha K_p (y_k - y^*_k)
\end{equation}

\begin{equation} \label{error_evolution1}
\left. \frac{\diff^{ } y}{\diff t^{ }} \right|_k - \left. \frac{\diff^{ } y^*}{\diff t^{ }}  \right|_k  = ( F_{k} - F_{k-1} ) + \alpha K_p (y_k - y^*_k)
\end{equation}

\noindent
Since the ultra-local model is valid only over a very short period of time, it is assumed that the variation of the estimation of $F$ is smooth according to the time i.e. the variation of $F_k$ is bounded by an $\varepsilon_M > 0$ such as:

\begin{equation} \label{bound_Fk}
|F_k - F_{k-1}| < \varepsilon_M
\end{equation}

\noindent
with $\varepsilon_M = 0.1$ as a convenient value. From this latter assumption and \eqref{error_evolution1}, it implies that the tracking error  converges towards zero provided a relatively small sampling period $h$. Notice that, as previously discussed, both $\alpha$ and $K_p$ are linked to the dynamic rate of the convergence. Denote $\varepsilon_k = y_k - y^*_k$ and $\dot{\varepsilon_k}$ its associated discrete time-derivative, we have:

\begin{equation}
\dot{\varepsilon}_k  = \varepsilon_M  + \alpha K_p {\varepsilon}_k 
\end{equation}

\begin{equation}\label{eq:dyn_epsilon}
\varepsilon_k = \frac{1}{ \alpha K_p } ( \dot{\varepsilon}_k - \varepsilon_M ) 
\end{equation}

The next step is to consider firstly the parameter $\alpha$ as time-varying and define the Lagrangian function, whose numerical solution of the corresponding optimization problem drives the evolution of $\alpha$ with respect to the tracking error in \eqref{eq:cont_1c}.

\subsection*{Discrete Euler-Lagrange equation}

Define now $\alpha(t)$ as a function of the time and define the Lagrangian function such as:
\begin{equation}\label{eq:int_cont}
 \mathcal{L(\alpha)} = \int_0^T L(s, \alpha(s), \dot{\alpha}(s)) \, \diff s
\end{equation}
\noindent
To discretize (\ref{eq:int_cont}), one consider the usual quadrature formulae such that:

\begin{equation}
 \int_{t_k}^{t_{k+1}} f(s) \diff s \approx (t_{k+1} - t_k) f( t_k)
\end{equation}
\noindent
where $t_{k+1} = t_k + h$ and $h > 0$ is the sampling time.

\noindent
The corresponding discrete Lagrangian function reads:

\begin{equation}\label{eq:int_disc}
 \mathcal{L(\alpha)} = \sum_{k = 0}^{N-1} (t_{k+1} - t_k) L(t_k, \alpha(k), \dot{\alpha}(k))
\end{equation}
\noindent
where $t_0 = 0, \, t_N = T, \, h = T/N$.

\noindent
The time-derivative of the function $\alpha$ is discretized and let us introduce the notations $\Delta^+$ and $\Delta^-$ associated respectively to the derivative taken at the instant $k$ and the instant $k-1$:

\begin{equation}
\left. \frac{\diff \alpha}{\diff t} \right|_{k} \approx  \frac{\alpha_{t_{k+1}} - \alpha_{t_k}}{h} := (\Delta^+ \alpha)_k
\end{equation}

\begin{equation}
\left. \frac{\diff \alpha}{\diff t} \right|_{k-1} \approx  \frac{\alpha_{t_{k}} - \alpha_{t_{k-1}}}{h} := (\Delta^- \alpha)_{k-1}
\end{equation}

\noindent
Hence:

\begin{equation}\label{eq:int_disc2}
 \mathcal{L(\alpha)} = \sum_{k = 0}^{N-1} (t_{k+1} - t_k) L \left( t_k, \alpha(k), \frac{\alpha_{t_{k+1}} - \alpha_{t_k}}{h} \right)
\end{equation}

\noindent
where $t_0 = 0, \, t_N = T, \, h = T/N$.

\noindent
From the well-known continuous Euler-Lagrange equation \cite{brunt}:

\begin{equation}
\frac{\diff}{\diff t} \left( \frac{\partial L}{\partial \dot{\alpha}} \right) = \frac{\partial L}{\partial \alpha}
\end{equation}

\noindent
the discrete Euler-Lagrange equation reads:

\begin{equation}
\begin{array}{c}
\displaystyle{\frac{1}{h} \left[ \frac{\partial L}{\partial \dot{\alpha}} (t_k, \alpha_k, (\Delta^+ \alpha)_k) - \frac{\partial L}{\partial \dot{\alpha}} (t_{k-1}, \alpha_{k-1}, (\Delta^- \alpha)_{k-1}) \right] = } \\
\displaystyle{ \frac{\partial L}{\partial {\alpha}} (t_k, \alpha_k, (\Delta^+ \alpha)_k) }
\end{array}
\end{equation}
\noindent
that is equivalent to:
\begin{equation}\label{eq:EL_disc}
{\Delta^- \left( \frac{\partial L}{\partial \dot{\alpha}} (t_k, \alpha_k, (\Delta^+ \alpha)_k) \right)_k =
\frac{\partial L}{\partial {\alpha}} (t_k, \alpha_k, (\Delta^+ \alpha)_k) }
\end{equation}

This discrete Euler-Lagrange equation will be coupled to the model-free control in the sequel, hence allowing to formulate the variational principle as an optimization problem.

\subsection*{Variational formulation}

The dynamic of the controlled error, that has been derived in  \eqref{error_evolution1}, provides a direct connexion between $\alpha$ and the convergence of the model-free controller \eqref{eq:dyn_epsilon}. To formulate the model-free controller as a Lagrangian, it is necessary to impose an additional dynamic to the parameter $\alpha$ by completing \eqref{eq:dyn_epsilon} with a extra time-derivative term.

Define $\gamma(t) = 1 / \alpha(t)$ (resp. the sample $\gamma_k$), eq. \eqref{eq:dyn_epsilon} can be rewritten as a coupled equation containing both error $\epsilon_k$ and $\gamma_k$ dynamics, such as:

\begin{equation}\label{eq:dyn_epsilon2}
\varepsilon_k = \frac{\gamma_k}{ K_p } \dot{\varepsilon}_k - \frac{\gamma_k}{ K_p } \varepsilon_M  + K_{\gamma}  \left. \frac{\diff \gamma}{\diff t} \right|_{k}
\end{equation}
where $K_{\gamma} \in \mathbb{R}^{*, +}$ is a coupling parameter. The derivative of $\gamma$ is written:

\begin{equation}
\left. \frac{\diff \gamma}{\diff t} \right|_{k} \approx  \frac{\gamma_{k} - \gamma_{k-1}}{h} 
\end{equation}

\noindent
The variational problem corresponds to the formulation of the minimization of the dynamic of the error, including the $\gamma_k$ dynamics \eqref{eq:dyn_epsilon2}. It is written (symbolically):

\begin{equation}\label{eq:dyn_epsilon3}
\begin{array}{c}
\displaystyle{\min_{\gamma_k} \int_0^t \varepsilon_k \, \diff \tau = } \\
\displaystyle{\min_{\gamma_k} \int_0^t \left\{ \frac{\gamma_k}{ K_p } \dot{\varepsilon}_k - \frac{\gamma_k}{ K_p } \varepsilon_M + K_{\gamma} \left. \frac{\diff \gamma}{\diff t} \right|_{k} \diff \tau \right\} }
\end{array}
\end{equation}
\noindent
Applying \eqref{eq:int_disc2}, the resulting discrete "model-free Lagrangian" is written:

\begin{equation}
\mathcal{L}_k(\gamma) = \sum_0^{N-1} \frac{\gamma_k}{ K_p } \dot{\varepsilon}_k - \frac{\gamma_k}{ K_p } \varepsilon_M  
+ K_{\gamma} \frac{\gamma_{t_{k+1}} - \gamma_{t_k}}{h}
\end{equation}
\noindent
where $t_0 = 0, \, t_N = T, \, h = T/N$.
Applying now the discrete Euler-Lagrange equation (\ref{eq:EL_disc}), we have:

\begin{equation}
\frac{\partial \mathcal{L}_k(\gamma)}{\partial \dot{\gamma}} = \frac{K_{\gamma}}{K_p} \dot{\gamma}
\end{equation}

\begin{equation}
 \frac{\partial \mathcal{L}_k(\gamma)}{\partial \gamma} = \frac{\dot{\varepsilon}_k}{K_p} - \frac{\varepsilon_M }{ K_p }  
\end{equation}

\begin{equation}
\Delta^- \left( \frac{ \partial \mathcal{L}_k}{\partial \dot{\gamma}} \right)(t_k) = \frac{\partial \mathcal{L}_k}{\partial {\gamma}} (t_k)
\end{equation}

\begin{equation}\label{eq:symp_int_final}
{ K_{\gamma} \frac{\gamma_{k+1} - 2 \gamma_k + \gamma_{k-1}}{h^2} = \dot{\varepsilon}_k - \varepsilon_M  }
\end{equation}
\noindent
The latter can be written as a difference equation: 

\begin{equation}\label{eq:symp_int_final2}
\gamma_{k+1} = \Gamma (\dot{\varepsilon}_k - \varepsilon_M) + 2 \gamma_k - \gamma_{k-1} 
\end{equation}
with $\Gamma = \frac{h^2}{K_{\gamma}}$. 
The symplectic integrator \eqref{eq:symp_int_final2} provides a relationship to describe the dynamic evolution of $\gamma$ coupled with the modified model-free law \eqref{eq:dyn_epsilon2}. Remark that \eqref{eq:symp_int_final2} is naturally unstable and thanks to the coupling with the control law, its behavior remains a priori bounded.

This concludes the development of the derivation of the symplectic integrator that minimizes the error evolution \eqref{eq:dyn_epsilon2}.

\subsection*{Symplectic-based model-free controller}

The proposed { symplectic-based model-free controller} is a combination of the discrete model-free controller (\ref{eq:cont_1c}) and the symplectic integrator:

\begin{equation}\label{eq:mfc_gamma_full}
\left\{ \begin{array}{c}
 \begin{array}{c}
\displaystyle{ u_{k} = u_{k-1} - \frac{1}{\alpha} \left\{\left. \frac{\diff^{ } y}{\diff t^{ }} \right|_{k-1} - 
\left. \frac{\diff^{ } y^*}{\diff t^{ }} \right|_k \right\} + } \\ 
\displaystyle{ K_p(y_k - y^*_k) + K_{\gamma} \left. \frac{\diff \gamma}{\diff t} \right|_{k}  }
\end{array} \\
\displaystyle{\gamma_{k+1} = \Gamma (\dot{\varepsilon}_k - \varepsilon_M ) + 2 \gamma_k - \gamma_{k-1} }\\
\end{array} \right.
\end{equation}
\noindent
where $(\varepsilon_M, K_p, K_{\gamma}, \alpha, \Gamma) \in \mathbb{R}^{5, *, +}$.

\noindent
Equation \eqref{eq:mfc_gamma_full} points out a strong coupling  between the model-free control law \eqref{eq:cont_1c} and the proposed variational integrator \eqref{eq:symp_int_final2}, whose purpose is to induce faster response to the tracking with respect to process changes and external disturbances.

\section*{IV. Numerical simulations}

\begin{figure}[h]
\centering
\includegraphics[width=3.2in]{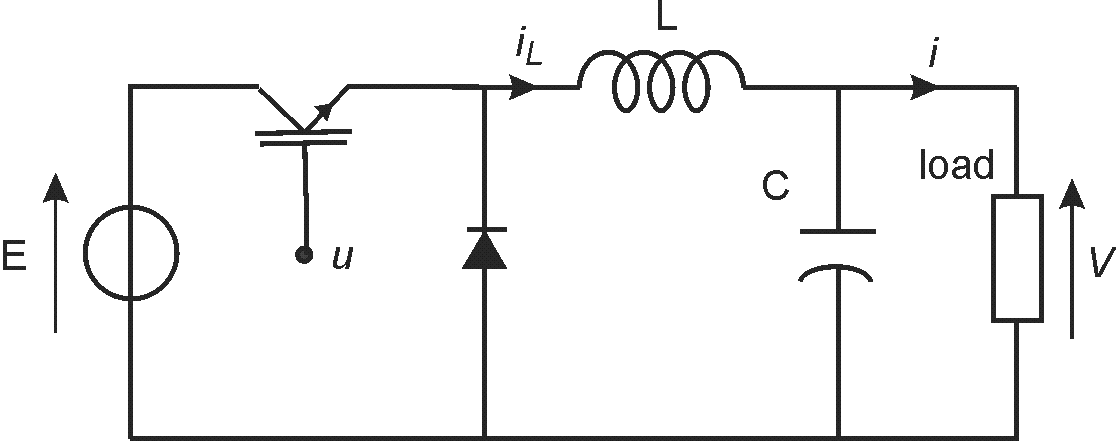}
\caption{Buck converter.}
\label{fig_buck}
\end{figure}

Consider the dc/dc buck converter\footnote{The application of the model-free control to the buck converter was the first application of the model-free control in power electronics \cite{COMPEL}.} depicted in Fig. \ref{fig_buck} where $u$ is the duty-cycle.

\noindent
The usual state-space model (in continuous mode) reads:

\begin{equation}
\dot{ \mathbf{x} } =
 \begin{pmatrix}
    0 & -\frac{1}{L} \\
   \frac{1}{C} & - \frac{1}{RC} \\
   \end{pmatrix}
   \mathbf{x}+
   \begin{pmatrix}
   \frac{E}{L} 
   \\ 0
   \end{pmatrix} u \\
   \label{avg_ss}
\end{equation}
\noindent
The nominal parameters are $E = 20 V$, $L = 1$ mH, $R = 10 \, \Omega$  and $C = 10$ $\mu$F. The model is simulated with a RK45 routine over a simulation horizon of 0.01 s with a time-step $h = 10^{-6}$ s.
A perturbation is added during the control operation that consists in modifying the circuit (typically a change of the RC load of the converter) inducing variations of the resistance $R = R + \Delta R$ and the capacitor $C = C + \Delta C$ at two instants $t = 4$ ms and $t = 6$ ms (the global variations are $2 \Delta R$ and $2 \Delta C$ for respectively the resistance and the capacitor at the end of the simulation).

The properties of the proposed variational-based model-free control is illustrated in case of voltage control and power control through several perturbations\footnote{Simulation code can be found in the repository:
\\
$<$\texttt{https://github.com/LoicMichelControl/MFCsymplectic}$>$.}.

\subsection{Voltage control}

% $\alpha = 10 . 10^{-3}$ and $K_p = 2$ -> OK
In the state-space \eqref{avg_ss}, the output voltage $v = y = x_2$ is controlled following an output reference $v^*$. The parameters of the control have been set to $\alpha = 10 . 10^{-3}$ and $K_p = 2$ to reach good tracking performances under the nominal parameters of the converter without perturbation\footnote{{Practically, $\alpha$ and $K_p$ can be tuned following e.g. the proposed procedure described in \cite{med_brigitte}.}}.

To study the influence of the variational integrator on the closed-loop behavior, two cases of perturbations will be investigated: firstly, the closed-loop remains stable when the perturbation occurs, and secondly, the closed-loop becomes unstable when the perturbation occurs. 
\\
\subsubsection{Stable perturbation case}

% $\Delta R = 1 \, \Omega$  $\Delta C = 1$ nF -> OK
The perturbation is set to $\Delta R = 1 \, \Omega$ and $\Delta C = 1$ nF.

\begin{figure}[h]
\centering
\includegraphics[scale = 0.055]{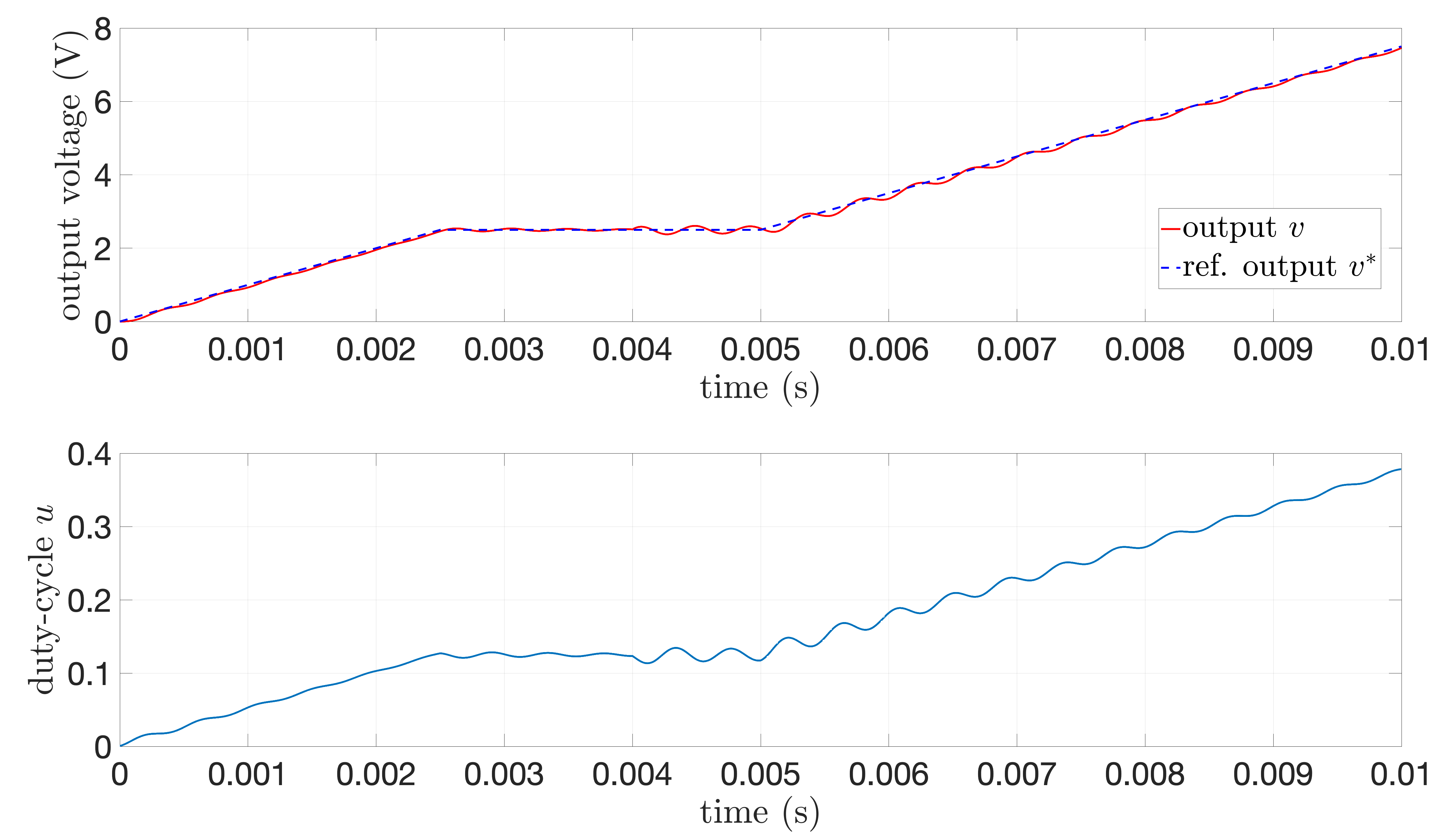}
\caption{Standard model-free control - controlled voltage $v$ and its reference $v^*$ (top) and associated duty-cycle $u$ (bottom) versus time (s).}
\label{buck_voltage_stand_1}
\end{figure}

% $\Gamma = 5.10^{-8}$ $\gamma_0 = 1$ -> OK
\noindent
Figure \ref{buck_voltage_stand_1} illustrates the tracking of a piecewise linear reference using the standard model-free control \eqref{eq:cont_1c}. 

\begin{figure}[h]
\centering
\includegraphics[scale = 0.055]{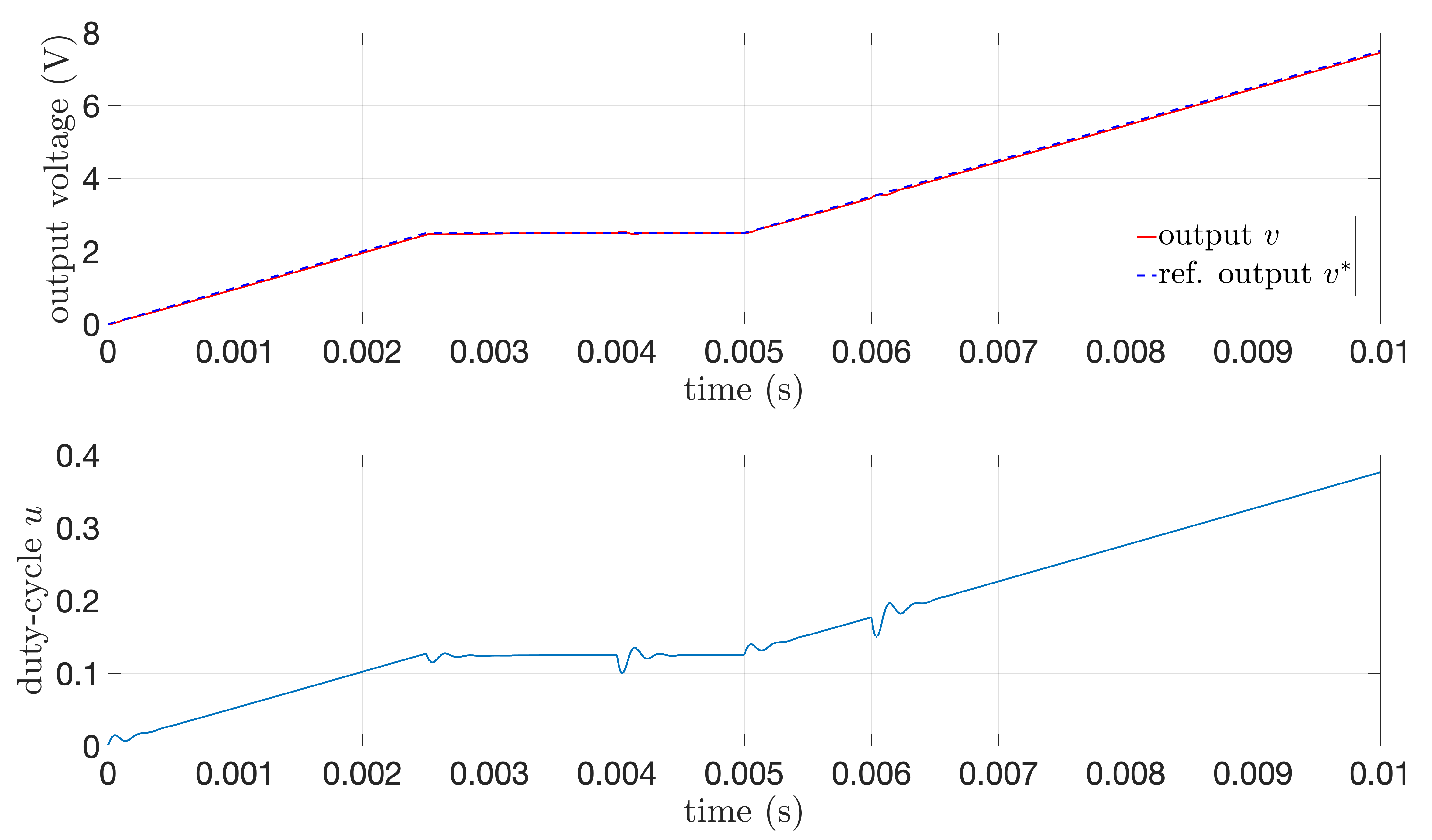}
\caption{Variational-based model-free control - controlled voltage $v$ and its reference $v^*$ (top) and associated duty-cycle $u$ (bottom) versus time (s).}
\label{buck_voltage_symp_2}
\end{figure}

\begin{figure}[h]
\centering
\includegraphics[scale = 0.05]{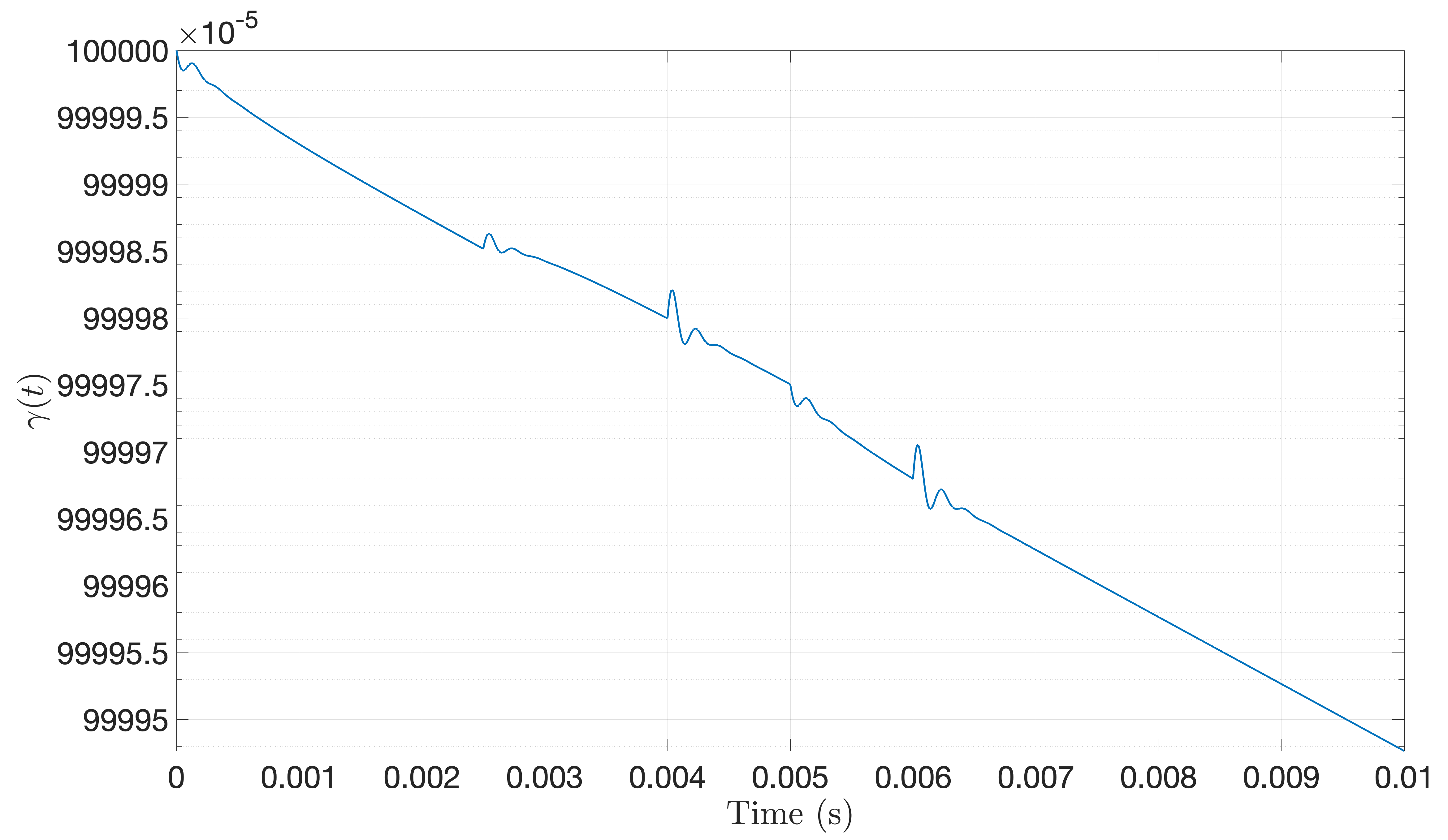}
\caption{Evolution of the parameter $\gamma$ versus time (s).}
\label{buck_voltage_gamma_3}
\end{figure}

\noindent
Figure \ref{buck_voltage_symp_2} shows the variational-based model-free control considering $\Gamma = 5.10^{-8}$, $K_{\gamma} = \gamma_0 = 1$ and Fig. \ref{buck_voltage_gamma_3} shows the evolution of the corresponding $\gamma(t)$ time-varying parameter.

\noindent
The tracking of the reference employing the symplectic-based control is much more smooth since the disturbance is greatly attenuated through the indirect action of $\gamma$ on the duty-cycle $u$: the time-varying parameter $\gamma$, which is affected by the variations of $v$ (due to the perturbation), adjusts accordingly the estimated $F$ in order to keep minimizing the tracking error (through the Lagrangian). In addition, the local changes of the slope of the output reference affect slightly the dynamic of $\gamma$.
\noindent
Remark that even small variations of $\gamma$ can have an impact on the closed-loop behavior since the variations of $\gamma$ are locally of order $4.10^{-6}$ when compensating the perturbation, compared with a global variation of $5.10^{-5}$ over the simulation horizon. This points out the fact that the disturbance rejection, and the effects of the changes of the output reference, are very correlated to the variations of $\gamma$. The behavior of $\gamma$ with respect to the controlled voltage $v$ and the reference $v^*$ emphasizes the strong coupling between the control law and the variational integrator, which tends to be a promising improvement to give more robustness to the tracking of time-varying systems and perturbation rejection.
\\
\subsubsection{Unstable perturbation case}
% $\Delta R = 5 \, \Omega$ -> OK
Keep the same control parameters and consider increasing the perturbation with $\Delta R = 5 \, \Omega$ and the same $\Delta C= 1$ nF. 

\begin{figure}[h]
\centering
\includegraphics[scale = 0.055]{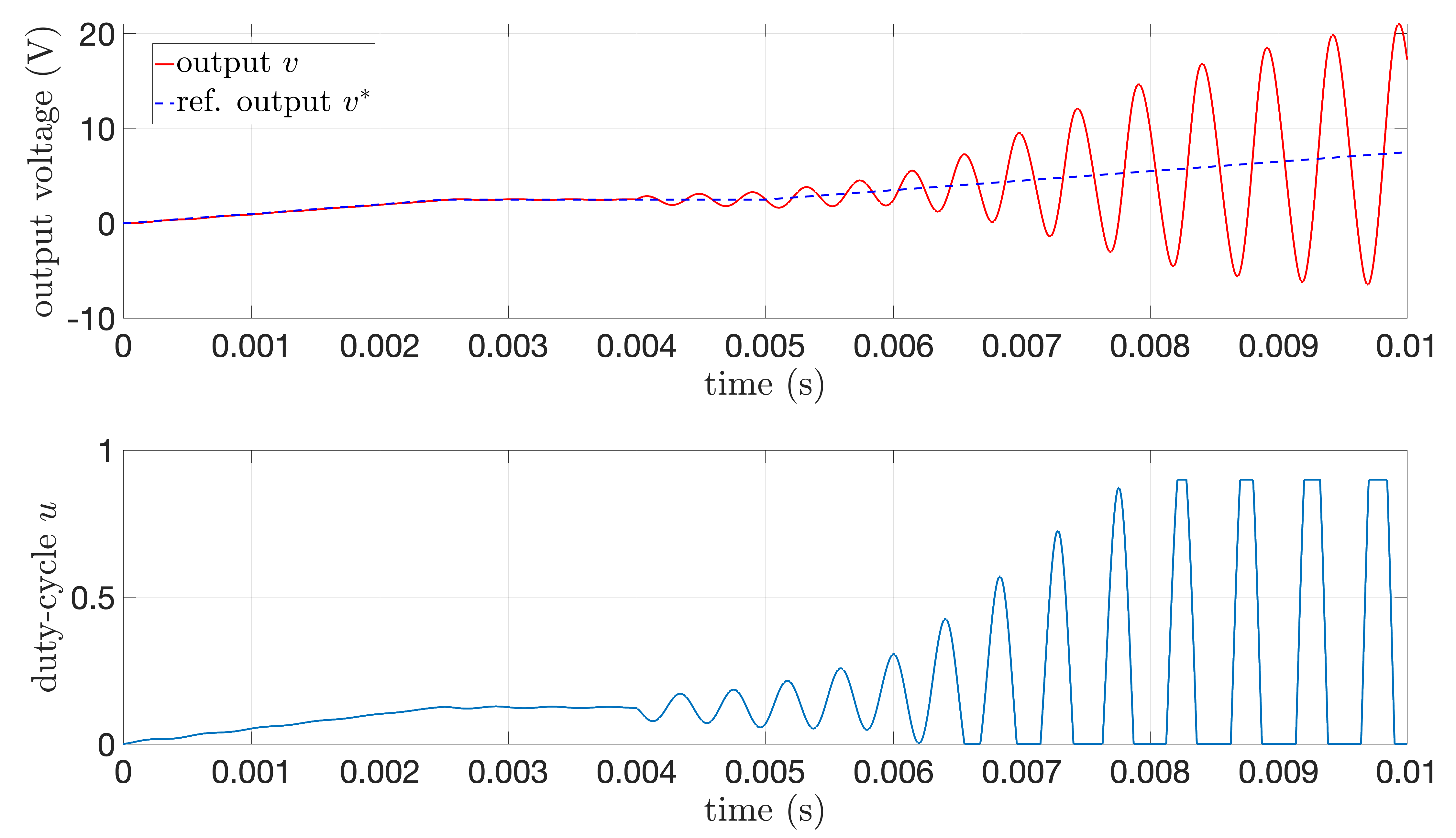}
\caption{Standard model-free control - controlled voltage $v$ and its reference $v^*$ (top) and associated duty-cycle $u$ (bottom) versus time (s) for a destabilizing perturbation.}
\label{buck_voltage_stand_unst_4}
\end{figure}

\noindent
Figure \ref{buck_voltage_stand_unst_4} shows that the closed loop becomes unstable because of the perturbation. 

\begin{figure}[h]
\centering
\includegraphics[scale = 0.055]{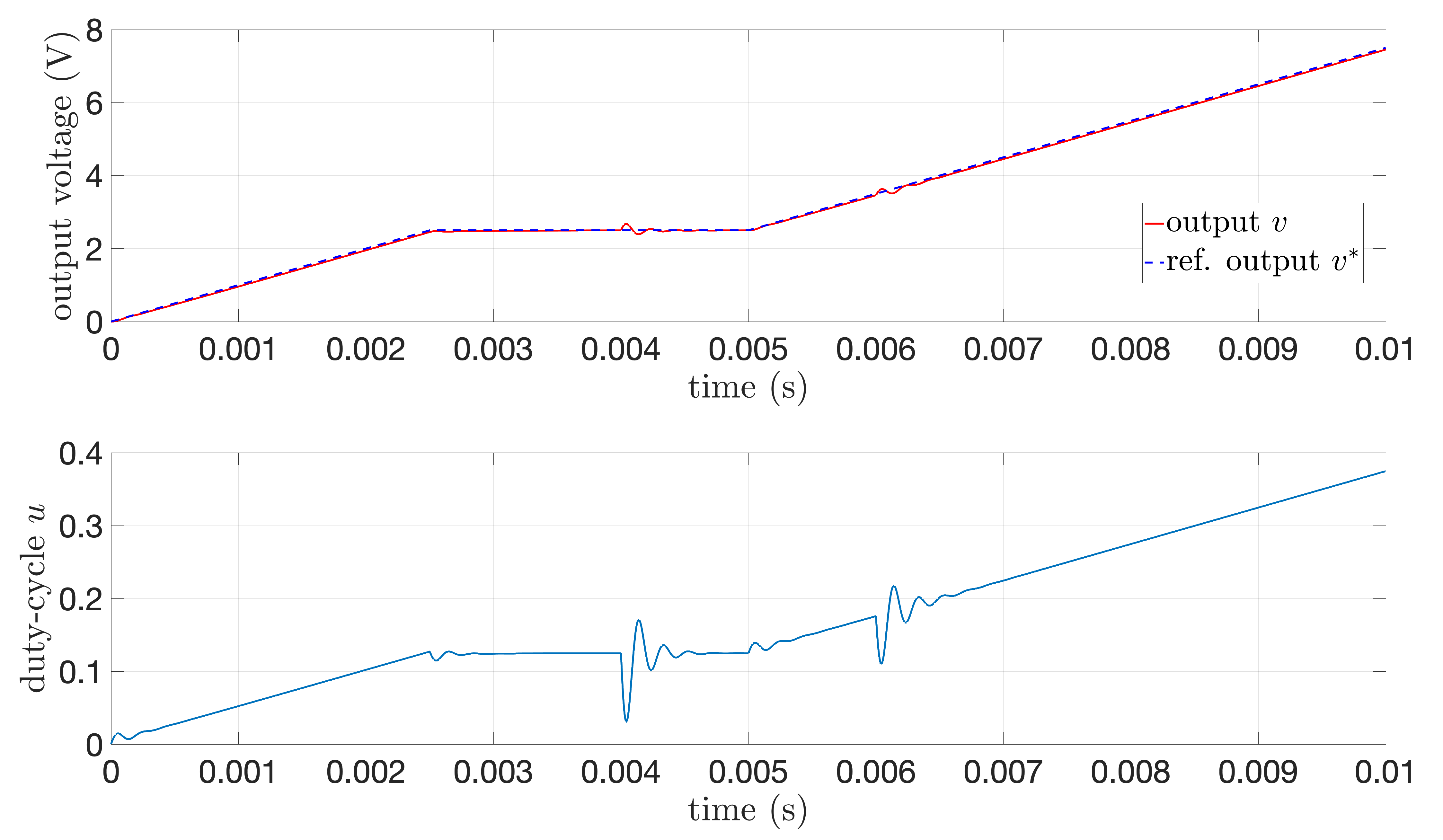}
\caption{Variational-based model-free control - controlled voltage $v$ and its reference $v^*$ (top) and associated duty-cycle $u$ (bottom) versus time (s) for a destabilizing perturbation.}
\label{buck_voltage_symp_unst_5}
\end{figure}

\noindent
However, the tracking of the reference employing the symplectic-based control remains stable (Fig. \ref{buck_voltage_symp_unst_5}). The comparison with Fig. \ref{buck_voltage_symp_2} shows that higher variations are induced on the duty-cycle $u$ to compensate the higher variations on $v$ induced by the perturbation. 

\subsection{Power control}

% $\alpha = 3 . 10^{-3}$ $K_p = 2$ -> OK

% $\Gamma = 5.10^{-8}$ $\gamma_0 = 3$ -> OK

% $\Delta R = 1 \, \Omega$ $\Delta C = 1$ nF -> OK

In the state-space \eqref{avg_ss}, the output power $p = y = x_1 x_2$ is controlled following an output reference $p^*$. The parameters of the control have been set to $\alpha = 3 . 10^{-3}$ and $K_p = 2$ to reach good tracking performances under the nominal parameters without perturbation.
The perturbation is set to $\Delta R = 1 \, \Omega$ and $\Delta C = 1$ nF.

\begin{figure}[h]
\centering
\includegraphics[scale = 0.055]{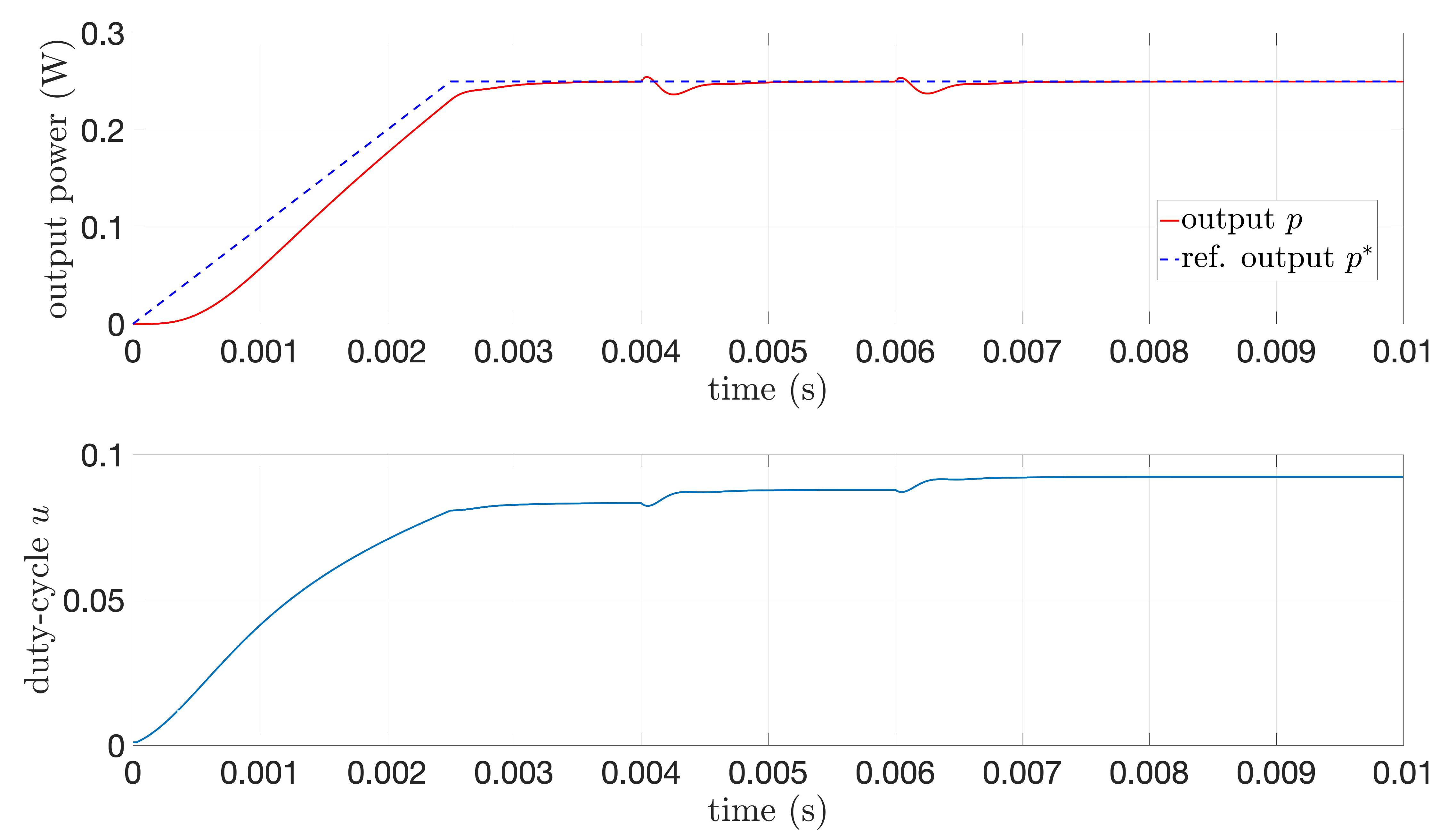}
\caption{Standard model-free control - controlled power $p$ and its reference $p^*$ (top) and associated duty-cycle $u$ (bottom) versus time (s).}
\label{buck_pwr_stand_6}
\end{figure}

\noindent
Figure \ref{buck_pwr_stand_6} illustrates the tracking of a piecewise linear reference using the standard model-free control \eqref{eq:cont_1c}. 

\begin{figure}[h]
\centering
\includegraphics[scale = 0.055]{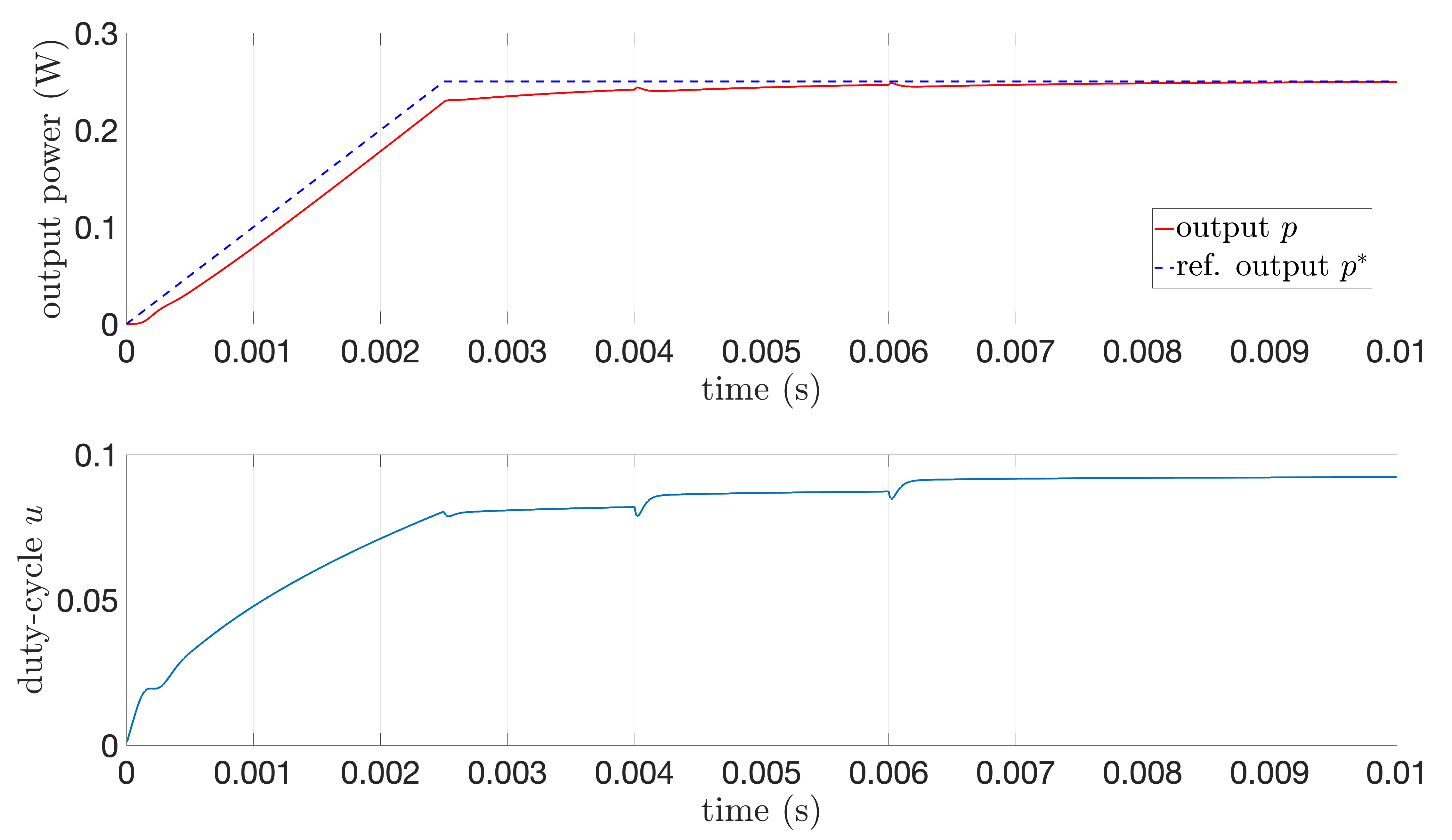}
\caption{Variational-based model-free control - controlled power $p$ and its reference $p^*$ (top) and associated duty-cycle $u$ (bottom) versus time (s).}
\label{buck_pwr_symp_7}
\end{figure}

\begin{figure}[h]
\centering
\includegraphics[scale = 0.05]{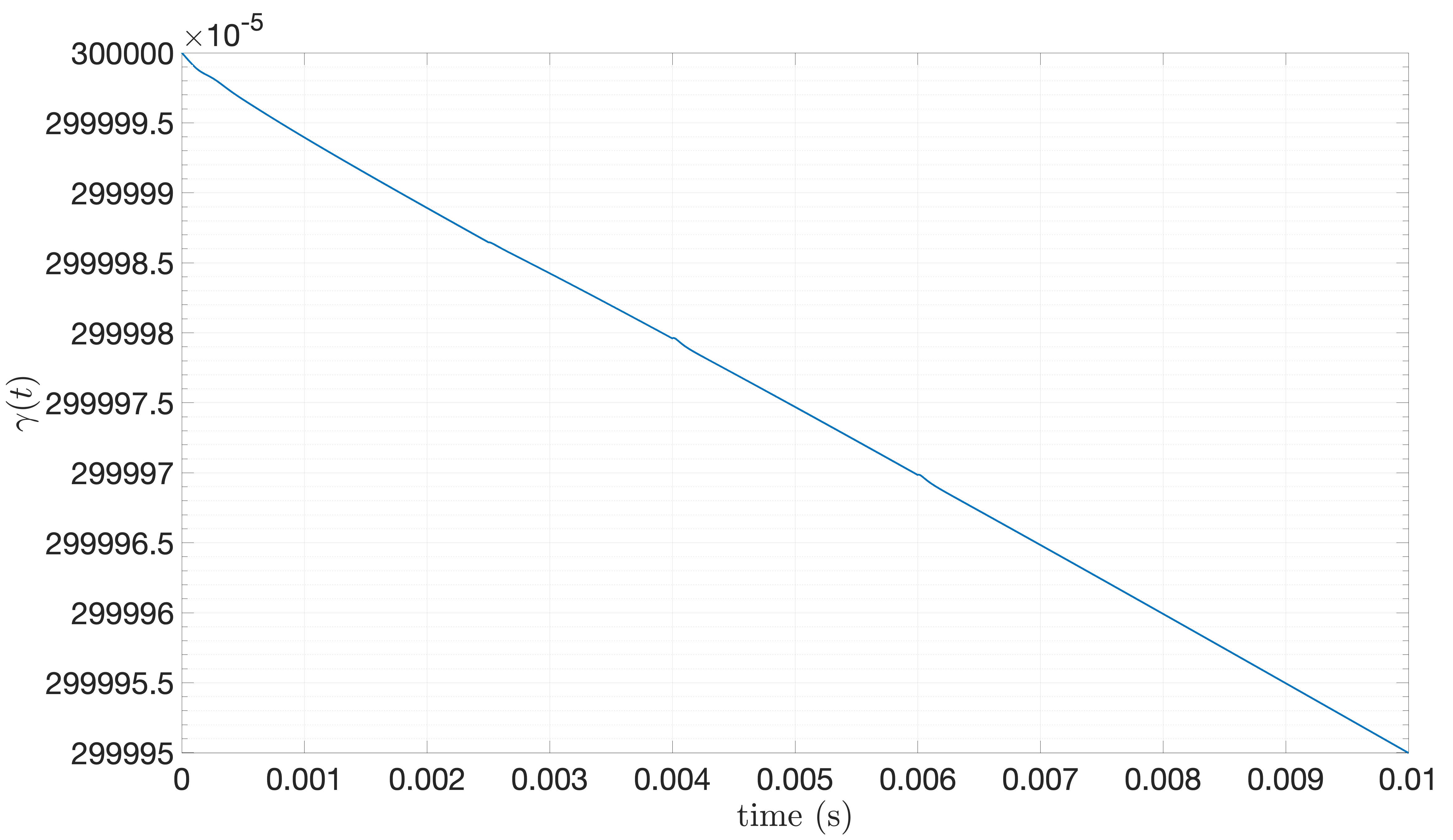}
\caption{Evolution of the parameter $\gamma$ versus time (s).}
\label{buck_pwr_gamma_8}
\end{figure}

\noindent
Figure \ref{buck_pwr_symp_7} shows the variational-based model-free control considering $\Gamma = 5.10^{-8}$, $K_{\gamma} = 1, \, \gamma_0 = 3$ and Fig. \ref{buck_pwr_gamma_8} shows the evolution of the corresponding $\gamma(t)$ time-varying parameter.

In this nonlinear case, the variational integrator shows promising performances toward the rejection of the perturbation. However, the dynamic of $\gamma$ may slow down the overall dynamic of the closed-loop to converge to the expected $p^*$ reference. Remark also that, like in the voltage control (see Figs. \ref{buck_voltage_symp_2} and \ref{buck_voltage_symp_unst_5}), the variations on the duty-cycle $u$ are more attenuated using the variational integrator than the standard control, meaning that, in a more general case, the proposed method would request less effort to the actuator to maintain the tracking of the reference.

\subsection{Parametrization study of the tracking error and $\gamma$}

\begin{figure}[h]
\centering
\includegraphics[scale = 0.05]{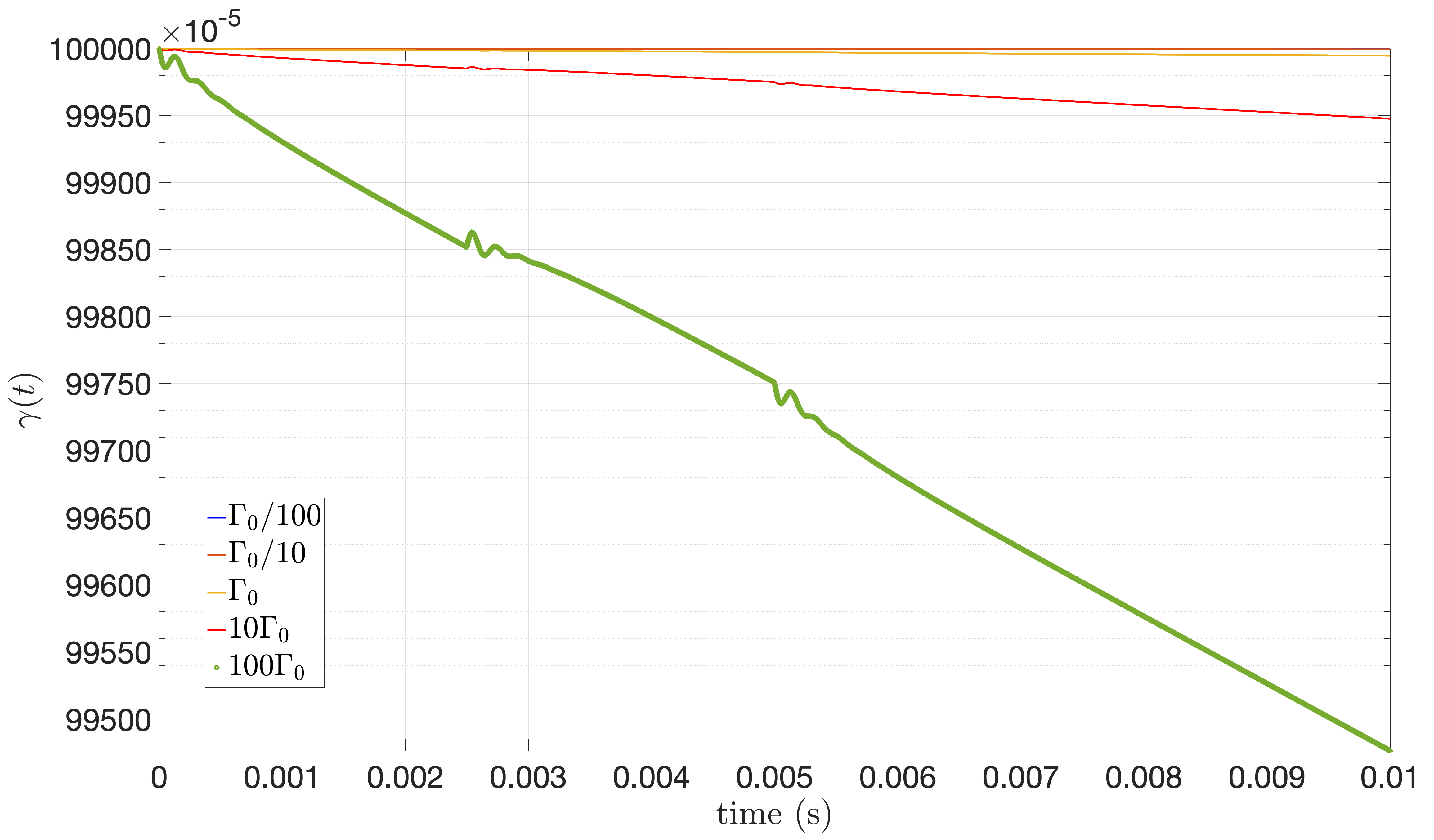}
\caption{Evolution of $\gamma$ versus time (s) with respect to several $\Gamma_0$.}
\label{parametric_gamma_Gamma0}
\end{figure}

\begin{figure}[h]
\centering
\includegraphics[scale = 0.05]{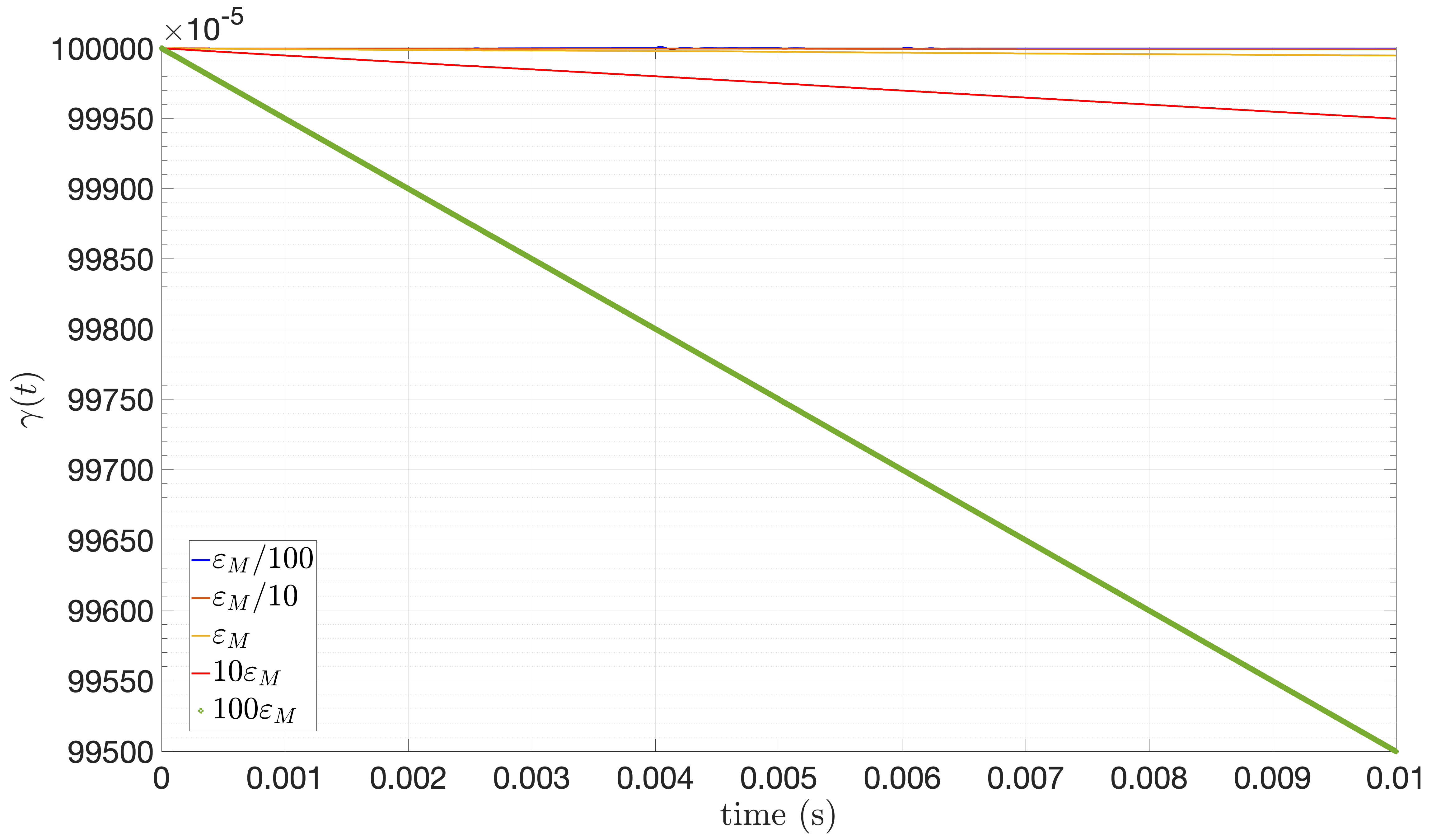}
\caption{Evolution of $\gamma$ versus time (s) with respect to several $\varepsilon_M$.}
\label{parametric_gamma_EpsilonM}
\end{figure}

\begin{figure}[h]
\centering
\includegraphics[scale = 0.05]{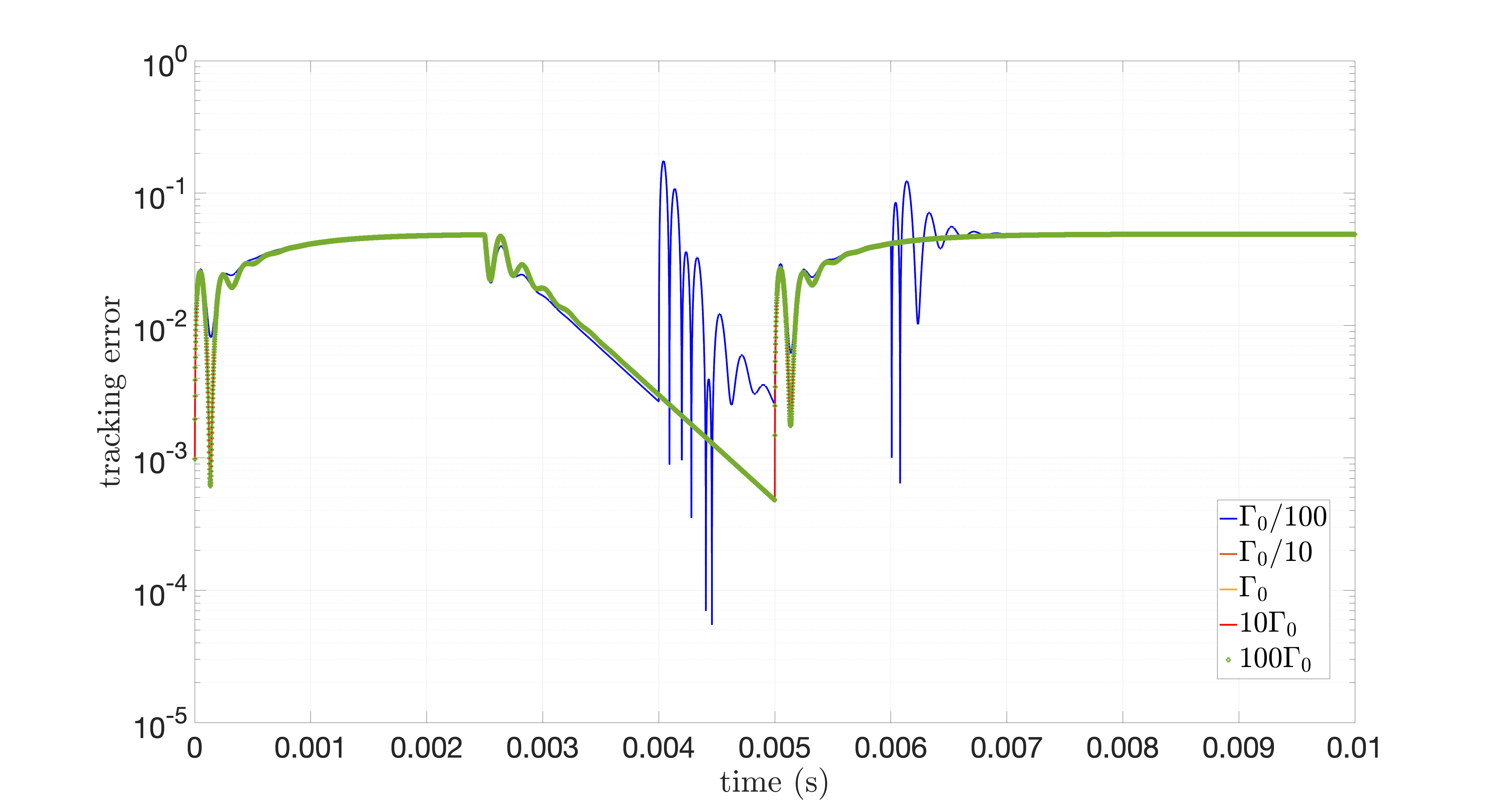}
\caption{Evolution of the tracking error (log scale) versus time (s) with respect to several $\Gamma_0$.}
\label{parametric_tracking_Gamma0}
\end{figure}

\begin{figure}[h]
\centering
\includegraphics[scale = 0.05]{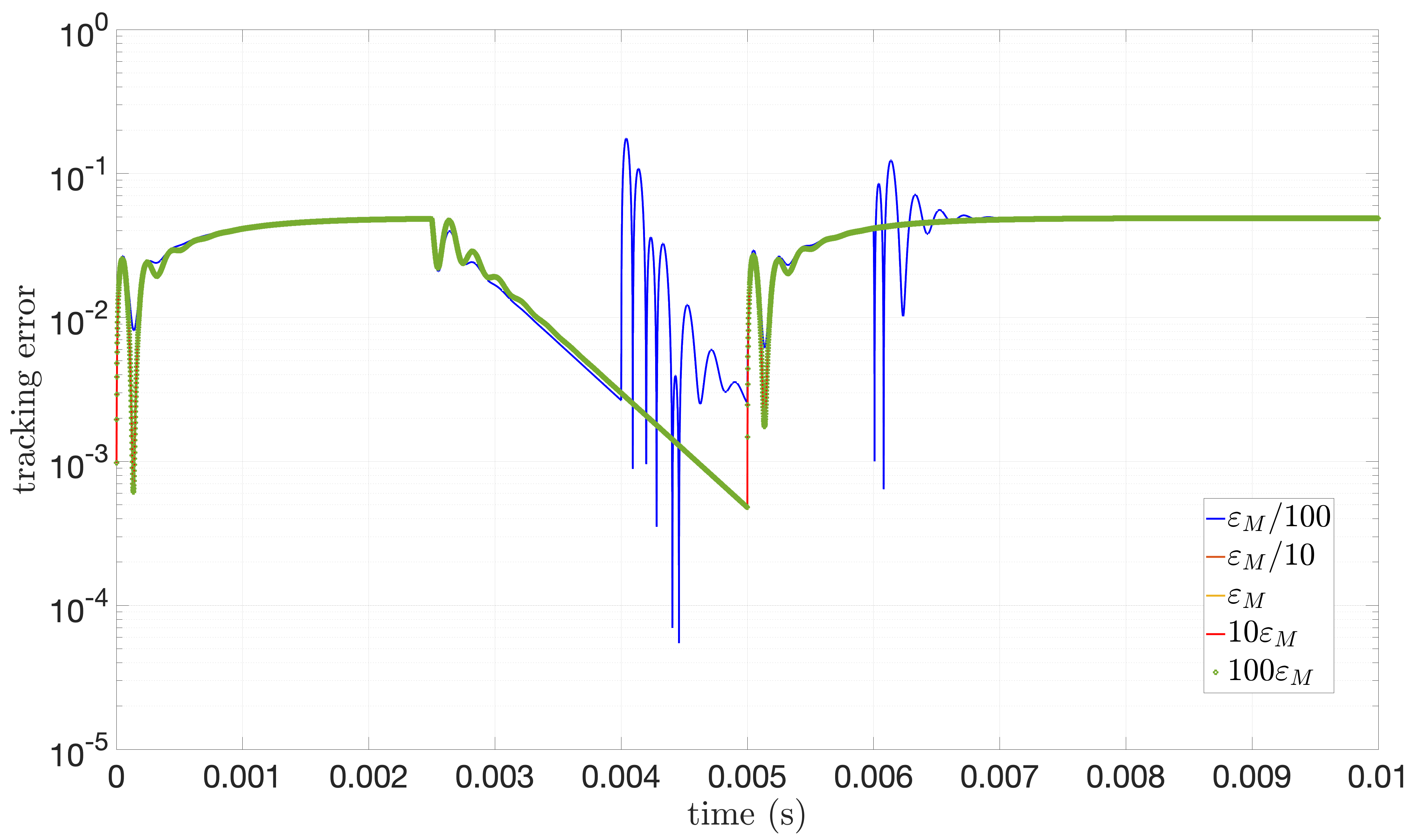}
\caption{Evolution of the tracking error (log scale) versus time (s) with respect to several $\varepsilon_M$.}
\label{parametric_tracking_EpsilonM}
\end{figure}

In this section, a parametrization study of the tracking error and the time-varying parameter $\gamma$ is performed in order to provide additional considerations about the choice of $\varepsilon_M$ (the bound associated to the variation of $F_k$ in \eqref{bound_Fk}) and the choice of $\Gamma$ with respect to the evolution of the tracking error and the evolution of $\gamma$ according to the time.
\noindent
Figures \ref{parametric_gamma_Gamma0} and \ref{parametric_gamma_EpsilonM} illustrate respectively the time-evolution of $\gamma$ considering several values of $\Gamma_0$ (with $\varepsilon_M$ = 0.1), and the time-evolution of $\gamma$ considering several values of $\varepsilon_M$ (with $\Gamma_0 = 5.10^{-8}$). 
\noindent
Figures \ref{parametric_tracking_Gamma0} and \ref{parametric_tracking_EpsilonM} illustrate respectively the time-evolution of the tracking error considering several values of $\Gamma_0$ (with $\varepsilon_M$ = 0.1), and the time-evolution of the tracking error considering several values of $\varepsilon_M$ (with $\Gamma_0 = 5.10^{-8}$).

\noindent
The parameter $\Gamma_0$ plays a significant role to determine the rate of variations of $\gamma$ as the evolution behaves with a linear tendency in all simulations (Fig. \ref{parametric_gamma_Gamma0}). Nevertheless, lower values may diminish the "reaction" to compensate the disturbances, which affects the tracking error and decreases the performances toward the perturbation rejection (Fig. \ref{parametric_tracking_Gamma0}). Higher values of $\Gamma_0$ does not affect significantly the tracking error and increase the "reaction" to compensate the disturbances, as well as increase the rate of variations of $\gamma$. This observation tends to enforce the fact that the sensitivity of the control toward the perturbation rejection 
is increased with  high local variations of $\gamma$. The parameter $\varepsilon_M$ plays a similar role since low values decrease significantly the tracking performances (Fig. \ref{parametric_tracking_EpsilonM}) and affect accordingly the variations of $\gamma$ (Fig. \ref{parametric_gamma_EpsilonM}). Note that a very low $\varepsilon_M$ would be restrictive regarding the evolution of $Fk$ in \eqref{bound_Fk}, hence compromising the integrity of the variational-based coupling with the model-free control. 

%A trade-off between $\varepsilon_M$ and $\Gamma_0$ would determine an interesting operating point that

\section*{V. Conclusion and perspectives}

This preliminary work presented a variational formulation of the Fliess's model-free control approach. The proposed control law is coupled with a symplectic integrator to auto-tune the key parameter of the controller. The perspectives include a complete study of the asymptotic stability of the proposed variational-based model-free control, including an analysis of the 
influence of the auto-tuning on the transient response efficiency.
The improvement of the strategy is also of strong interest considering, for instance, the investigation of multi-rate integration  \cite{variationalmultirateintegrators} to allow the auto-tuning on different time-scales.
Furthermore, the experimental validation of the approach on complex systems involving switching conditions (i.e. hybrid dynamical systems), like complex topologies of power converters and multiple controlled tasks management in the robotic gripper framework \cite{gripper}. In addition, such symplectic framework can be of interest to be transposed to other advanced control laws in order to attempt improving their robustness.

%\textcolor{red}{Je pense avoir fait une erreur dans la formule initiale de l'int\'egrateur de mon papier arxiv (suite \`a la difficult\'e d'interpr\'etation de la notation Marsden-West). Il n'y a donc pas {\it a priori} de second terme $\gamma_k$.}

% \textcolor{red}{Sauf erreur, $\nu$ multiplie $K_{\gamma}$ apr\`es d\'erivation $D_1$. Est-il n\'ecessaire finalement de conserver la puissance $\nu$ (prise \`a la valeur 2) si le coefficient $K_{\gamma}$ multiplie l'ensemble (et compense $\nu$) ?  Sauf erreur, les modifications des param\`etres $\nu$ et $K_{\gamma}$ ne modifient pas la r\'eponse en r\'egulation pour le m\^eme cas d'\'etude. \underline{Mais}, l'adjonction arbitraire du terme $\varepsilon_M$ (Fig. 6) tend \`a annuler les oscillations de $\gamma$ en conservant la pr\'ecision d'optimisation sur $y$...}

% \textcolor{blue}{Questions \`a traiter : stabilit\'e globale; comment changer la pseudo-p\'eriode de l'int\'egrateur; possibilit\'e de "r\'eglage" de la finesse de l'int\'egrateur $(\forall t > 0, \varepsilon(t) \rightarrow 0 )$; application aux cas non-lin\'eaires et syst\`emes commut\'es. Possibilit\'e de quantifier de mani\`ere explicite l'efficacit\'e de l'optimisation (\'evaluation explicite de $\int \varepsilon$).}

\section*{ACKNOWLEDGMENT}

The author is thankful to Prof. J. Cresson for the useful discussions regarding the development of the symplectic framework, and is grateful to Prof. J.-P. Barbot for his useful comments and support to prepare the paper.

\bibliographystyle{IEEEtran}
\bibliography{CCTA_Symplectic_MFC.bib}

%%%%%%%%%%%%%%%%%%%%%%%%%%%%%%%%%%%%%%%%%%%%%%%%%%%%%%%%%%%%%%%%%%%%%%%%%%%%%%%%

\end{document}